# Zero Dimensional Polariton Laser in a Sub-Wavelength Grating Based Vertical Microcavity

Bo Zhang,[1] Zhaorong Wang,[1] Sebastian Brodbeck,[2] Christian Schneider,[2] Martin Kamp,[2] Sven Höfling,[2,3] and Hui Deng[1]

[1]Department of Physics, University of Michigan, Ann Arbor, Michigan, 48109

[2]Technische Physik, Physikalisches Institut and Wilhelm Conrad Roentgen Research Center for Complex Material Systems, University of Wuerzburg, D-97074 Wuerzburg, Germany

[3]SUPA, School of Physics and Astronomy, University of St Andrews, St Andrews, KY16 9SS, United Kingdom

## Abstract

Semiconductor exciton-polaritons in planar microcavities form coherent two-dimensional condensates in non-equilibrium. However, coupling of multiple lower-dimensional polariton quantum systems, critically needed for polaritonic quantum device applications and novel cavity-lattice physics, has been limited due to the conventional cavity structures. Here we demonstrate full confinement of the polaritons non-destructively using a hybrid cavity made of a single-layer sub-wavelength grating mirror and a distributed Bragg reflector. Single-mode polariton lasing was observed at a chosen polarization. Incorporation of a designable slab mirror into the conventional vertical cavity, when operating in the strong-coupling regime, enables confinement, control and coupling of polariton gasses in a scalable fashion. It may open a door to experimental implementation of polariton-based quantum photonic devices and coupled cavity quantum electrodynamics systems.

**INTRODUCTION**

Semiconductor microcavity exciton-polaritons [1] have recently emerged as a unique, open system for studying non-equilibrium quantum orders [2–4]. Exciton-polaritons are formed via strong coupling between the excitons and photons. Due to the excitonic component, polaritons are massive, weakly interacting quasi-particles that feature strong nonlinearity and rich manybody physics [5]. Due to the mixing with the photon, polaritons have an effective mass $10^{-8}$ of the Hydrogen atom mass, and they are relatively insensitive to disorders or localization potentials in the active media. Hence polaritons exhibit quantum coherence over macroscopic scales at high critical temperatures. Polaritons in quantum-well (QW) microcavities couple out of the cavity at a fixed rate while conserving the energy and in-plane wavenumber, providing direct experimental access unavailable in typical manybody quantum systems. Hallmarks of non-equilibrium condensation and superfluidity have been widely observed in isolated two-dimensional polariton systems (Ref. [3] and references therein).

The foundational work on 2D polariton systems has inspired theoretical schemes for polariton-based quantum circuits [6–8], quantum light sources [9–12], and novel quantum phases [4]. Experimental implementation of these schemes requires control, confinement and coupling of polariton systems, which remain challenging with the conventional microcavity structure. Important features of a versatile experimental platform based on polaritons include: firstly, well defined 0D polaritons as building blocks of a coupled system, secondly, the establishment (i.e., survival) of non-equilibrium quantum phase in each 0D polariton cell typically manifested as polariton lasing, thirdly, controllable coupling among the 0D cells, and lastly, individual addressability and control of each cell. In conventional polariton-cavities, the thick mirrors, made of distributed Bragg reflectors (DBRs), make it difficult to confine or control the polaritons beyond the perturbative regime. Most existing methods to control the polaritons lead to a weak modulation potential that modifies the system's properties without reducing its dimensionality from 2D to 0D. Examples include weakly confining the excitons via mechanical strain [13] and periodic modulation of the optical modes via surface patterning [14, 15]. Advanced techniques have been developed to embed apertures inside the cavity [16, 17], which has created 0D polariton cells but polariton lasing has not been reported so far. Alternatively, 0D polariton systems were also created via direct etching of the vertical cavity into pillars [18–21]. Using this method, two groups have achieved polariton lasing in the pillars recently [22–24], and thus satisfying the first two requirements. However, this approach requires a destructive plasma etching throughout the 4-6 μm tall cavity structure as well as the active media layers, which excludes coupling between separate pillars. It is also unclear if further control of the polariton modes in each pillar would be possible. In this work, we demonstrate a polariton system in an unconventional cavity that could fill all the four requirements. The new cavity structure replaces the top DBR with a slab photonic crystal (PC) (as shown in Fig. 1), which enables confinement and control of the polariton modes by design [25–27]. At the same time, there is no destructive interface in the active media layers or the main cavity layers, hence coupling among multiple low-dimensional polariton cells is unhindered. We demonstrate with the new cavity system zero-dimensional polariton lasing at a chosen polarization.

**MATERIALS AND METHODS**

A schematic of our hybrid cavity polariton device is shown in Fig. 1(a). To fabricate the device, we first grew the planar structure by molecular beam epitaxy (MBE) on GaAs substrate, consisting of 30-pairs of bottom DBR, an AlAs λ/2 cavity layer, a 2.5 pairs of top DBR of $Al_{0.15}GaAs/AlAs$, and an $Al_{0.85}GaAs$ sacrificial layer followed by a $Al_{0.15}GaAs$ top layer. There are 12 GaAs QWs distributed at the three central anti-nodes of the cavity. We create square gratings of 5-8 μm in length (Fig. 1(b)) on the top layer via electron-beam lithography followed by a short plasma etching. A hydrochloric acid chemical etching was then applied to remove the sacrificial layer, followed by critical point drying. The fabricated gratings are ~80 nm thick, with a period of ~520 nm and a duty cycle of ~40%, and suspended on an air gap of ~300 nm. It is optimized as a high reflectance mirror for light polarized along the grating bar direction (TE-polarization). Figure 1(b) shows the scanning electron microscopy (SEM) image of the top-view of one of the devices.

Optical measurements were performed to characterize the properties of the cavity system. The sample was kept at 10-90 K in a continuous flow liquid-helium cryostat. A pulsed Ti-Sapphire laser at 740 nm was used as the excitation laser, with a 80 MHz repetition rate and 100 fs pulse duration. It is focused to a spot size of ~2 μm in diameter on the device from the normal direction with an objective lens of a numerical aperture of 0.55. The photoluminescence signal was collected with the same objective lens, followed by real space or Fourier space imaging optics, and sent to a 0.5 m spectrometer with an attached nitrogen cooled charge coupled device (CCD). The spectrally resolved real space and Fourier space distributions were measured by selecting a strip across the center of the Fourier space and real space distributions using the spectrometer's entrance slit. The resolution of the measurements was limited by the CCD pixel size to 0.3 /μ for Fourier space imaging and by the diffraction limit to 0.4 μm for real space imaging.

**RESULTS AND DISCUSSION**

Strong-coupling between excitons and TE-cavity modes were evident in the momentum space images of the emission from within the cavity, as shown in Fig. 2(a). Discrete lower polariton (LP) modes and a faint upper polariton (UP) branch were observed below and above the exciton energy, respectively, with dispersions distinct from that of the cavity photon (the red solid line). In contrast, the emission from outside the hybrid cavity region shows a flat, broad exciton band at the heavy hole exciton energy of $E_{exc}$ =1.551 eV (Fig. 2(b)). The energies of the polariton modes can be described as follows in the rotating wave approximation:

$$E_{UP,LP}(k) = \frac{1}{2}[E_{exc}(k) + E_{cav}(k) \pm \sqrt{(E_{exc}(k) - E_{cav}(k))^2 + 4\hbar^2\Omega^2}]. \quad (1)$$

Here $k$ is the inplane wavenumber, $E_{cav}$ is the un-coupled cavity energy and $2\hbar\Omega$ is the exciton-photon coupling strength, corresponding to the LP-UP splitting at the zero exciton-photon detuning. Using Eq. 1 and the measured values $E_{exc}(k=0)$ =1.551 eV, $E_{LP}(k=0)$ = 1.543 eV, and $E_{UP}(k=0)$ = 1.556 eV, we obtain $E_{cav}(k=0)$ = 1.548 eV and $2\hbar\Omega$ = 12 meV.

The discrete LP modes shows full three dimensional confinement of the polaritons. The lateral size of the hybrid cavity is determined by the size of the high index SWG. Outside the SWG, there is no cavity resonance and excitons are the eigen-excitations. Inside the SWG region, TE-polarized cavity modes strongly couple to excitons, leading to laterally confined TE-polarized polariton modes. TM-polarized excitons remain in the weak coupling regime. Since there is not a sharp lateral boundary where the cavity mode disappear, we model the effective confinement potential as an infinite Harmonic potential phenomenologically. The calculated energies of the LP modes are indicated by the dashed lines in Fig. 2(a), which agree very well with the measured LP resonances. For comparison, the confined cavity modes (crosses) and corresponding 2D dispersions of the LP, UP and cavity modes are also shown (solid lines).

The spatial profile of the confined LP modes are also measured via spectrally resolved real-space imaging, as shown in Fig. 2(c). The four lowest LP modes are well confined within the SWG region, while higher excited states spread outside and form a continuous band. The variances of the k-space and x-space wavefunctions along the detected direction are $\Delta k$ =0.85 /μm and $\Delta x$ =1.01 μm. Their product is $\Delta x \times \Delta k$ =0.86. It is slightly larger than the uncertainty limit of 0.5, which may be due to the diffusion of the LPs.

The absorption spectra of the modes were obtained via reflectance measurements. The spectrum measured normal to the sample (Fig. 2(d)) shows the three symmetric modes with the lowest mean inplane wavenumber: the UP ground state, the LP ground state, and the LP second excited states. Other polariton states have too small a spectral weight to be measured in reflectance. When measured at 3.5° from the sample normal, the 1st excited state of LPs was also observed (Fig. 2(e)).

A further confirmation of the strong-coupling regime is the temperature tuning of the resonances, as shown in Fig. 2(f). As the temperature increased, the LP and UP ground state energies red shifted and were measured via k-space PL. The exciton energy was measured directly in the planar region outside the SWG. The shift of the cavity photon energy was obtained from the shift of the 1st low-energy side minimum of the stopband. Anti-crossing of the LP and UP modes is evident. From the LP, exciton and cavity energies, we obtain the coupling strength $2\hbar\Omega(T)$ ~10 meV from 10 K to 80 K, showing that strong coupling persists to the liquid nitrogen temperature or higher.

Unlike planar DBRs, the grating breaks the in-plane rotational symmetry. As a result, the SWG mirrors can have high polarization selectivity. We optimized our SWG to have high reflectance for the TE mode and low reflectance for the orthogonal TM mode. Correspondingly, the polaritons are TE polarized, while the TM polarized excitons remain in the weak coupling regime. Figure 3 shows the PL intensity vs. the angle of linear polarization for the LPs and excitons at k ~ 0 within the SWG region, normalized by the maximum intensity. We fit the data with $I = A\cos(\theta - \phi)^2 + B$, where the fitting parameters $\phi$ depends on the orientation of the device, $A$ corresponds to linearly polarized light, and $B$ corresponds to a non-polarized background. Correspondingly, the degree of linear polarization is DOP= $(I_{max} - I_{min})/\{I_{max} + I_{min}) = A/(A + 2B)$. We obtained $A_{LP}$ =1.04 ± 0.04, $B_{LP}$ =0.05 ± 0.01, $\phi_{LP}$ = 71°± 1°, and DOP = 91.9% for the LPs, confirming that LPs are highly TE-polarized. For the excitons, we obtained $A_{exc}$ =0.891±0.001, $B_{exc}$ =0.0081± 0.0002, $\phi_{exc}$ = 161°±1° = $\phi_{LP}$+90°, and DOP = 98.2%, showing that excitons are polarized orthogonal to the LPs. Such control of the polariton polarization has not been possible with conventional DBR-DBR cavities and is unique to the SWG-based cavity.

Finally, we show that polariton lasing was achieved in the 0D hybrid cavity. As shown in Fig. 4 (a), the emission intensity I from the LP ground state increases sharply with the excitation power P at a threshold of $P_{th}$ =~ 5 KW/cm$^2$, characteristic of the onset of lasing. Interestingly, the emission intensity I varies with P quadratically both below and well above the threshold, except at very low excitation densities. This may be because the energy separation between the discrete modes is larger than $k_B T$ ~ 0.8 meV. As a result, relaxation to the ground state through LP-phonon scattering is suppressed compared to LP-LP scattering. Accompanying the transition, a sharp decrease of the LP ground state linewidth was measured. The minimum linewidth of 0.24 meV may be limited mainly by the intensity fluctuation of the pulsed excitation laser [28]. The LP energy increased continuously with the excitation density due to exciton-exciton interactions. The blueshift shows a linear dependence below threshold, it is suppressed near threshold, and shows a logarithmic dependence above threshold [22, 29]. The discrete energy levels are maintained across the threshold and remain distinctly below the uncoupled cavity energy.

The establishment of polariton lasing cofirms the quality of the 0D-polariton system. The threshold density is smaller or comparable to those measured in DBR-DBR pillar cavities [22, 23]. The linewidth deduction and blueshift are all within an order of magnitude difference compared to reported values in DBR-DBR planar or pillar microcavities [1, 22, 23]. Unlike DBR-DBR cavities, however, the polariton lasing we demonstrated takes place in *a priori* defined polarization, independent of excitation conditions.

**CONCLUSIONS**

In conclusion, we have demonstrated the first hybrid cavity incorporating a slab PC mirror, operating in the strong coupling regime, supporting polariton lasing. Three dimensional confinement of the polaritons was achieved by using a finite size SWG, with the QWs and the main cavity layers untouched. Polariton lasing in the ground state was readily observed.

Unique to the hybrid SWG-cavity, the LP is linearly polarized, while the orthogonally polarized exciton mode remains in the weak-coupling regime. The PL of the weakly-coupled

TM excitons provides direct access to the TE exciton reservoir that has not been available in conventional III-As cavities. It enables polarized polariton lasers [30–34] and simplifies quantum photonic devices based on single-spin polaritons [10–12, 35].

The integration of a slab PC mirror in a polariton system adds the flexibility to control the fundamental properties of polaritons by design, including the dimensionality and polarization, as demonstrated in this work. Exploring different PC designs will allow further modification of the properties of the polaritons. Importantly, the control of the polariton system is achieved without creating destructive interfaces in the active media or main cavity layer. Hence extension of the single 0D polariton system into multiple closed placed ones would allow the creation of controllably coupled polariton systems, while each 0D-cell in the coupled system can be separately controlled and probed. The demonstrated hybrid-cavity polariton system may provide a scalable architecture for the experimental implementation of coupled lattice cavity systems [4, 36].

**ACKNOWLEDGMENTS**

BZ, ZW, and HD acknowledge the support by the National Science Foundation (NSF) under Awards DMR 1150593 for measurements and OISE 1132725 for travel expenses, and the Air Force Office of Scientific Research under Awards FA9550 -12-1-0256 for device fabrication and characterization. CS, SB, MK and SH acknowledge the support by the State of Bavaria, Germany. The fabrication of the SWG was performed in the Lurie Nanofabrication Facility (LNF) which is part of the NSF NNIN network.

**FIGURE CAPTION**

FIG. 1. Examples of the hybrid cavity. (a) A schematic of a 0D hybrid cavity with a SWG mirror. (b) The top-view SEM image of a fabricated 0D cavity with a SWG of 5 μm×5 μm in size.

FIG. 2. Spectral properties of a 0D-polariton device. (a) The spectrally resolved momentum space image of the PL from a 0D cavity, which shows discrete LP modes and an UP mode. To show clearly the UP mode, the intensity of the upper panel is magnified by 40× compared to the lower panel. The straight red line at 1.551 eV corresponds to the independently measured exciton energy. The other solid lines are the calculated dispersions of the LP, UP and un-coupled cavity. The white dashed lines and the crosses (X) mark the position of the calculated discrete LP and cavity energies, respectively. (b) The spectrally resolved momentum space images of the exciton PL, measured from the un-processed part next to the SWG-DBR cavity. (c) The spectrally resolved real space image of the PL from the 0D cavity, showing the spatial profile of the discrete LP modes. (d)-(e) Reflectance spectra of the 0D cavity measured from (d) normal direction and
(e) from 3.5° from normal direction, both with an angular resolution of 0.27°. (f) Temperature dependence of the LP (stars), exciton (squares), and cavity resonances (circles).

FIG. 3. Polarizations of the polaritons and excitons in the hybrid-cavity polariton system. (a) Polar plots of the LP ground state intensity as a function of the angle of the linear polarization analyzer. The symbols are the data. The solid lines are fits using I=Acos$(θ−ϕ)^2$+B as described in the text, with a corresponding fitted linear degree of polarization of 91.9%. (b) Polar plot for the exciton emission intensity from within the SWG, corresponding to a fitted linear degree of polarization of 98.2% and orthogonal polarization compared to (a).

FIG. 4. Lasing properties of the 0D-polaritons. (a) Integrated intensity, (b) linewidth and (c) corresponding energy blue shift of the LP ground state v.s. the excitation density. The dashed lines in (a) are a comparison with quadratic dependence. The dashed lines in (c) are comparisons with the linear dependence below threshold and logarithmic dependence above threshold, respectively.

Figure 1

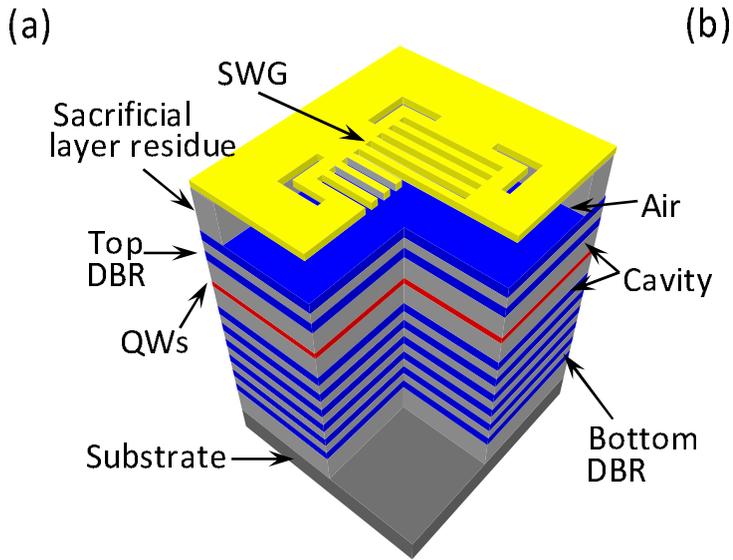 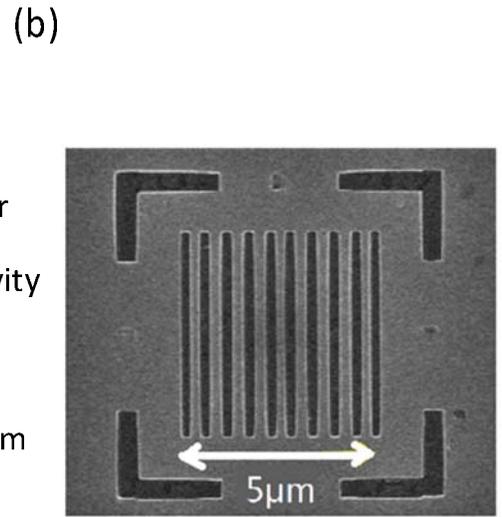

(a) (b)

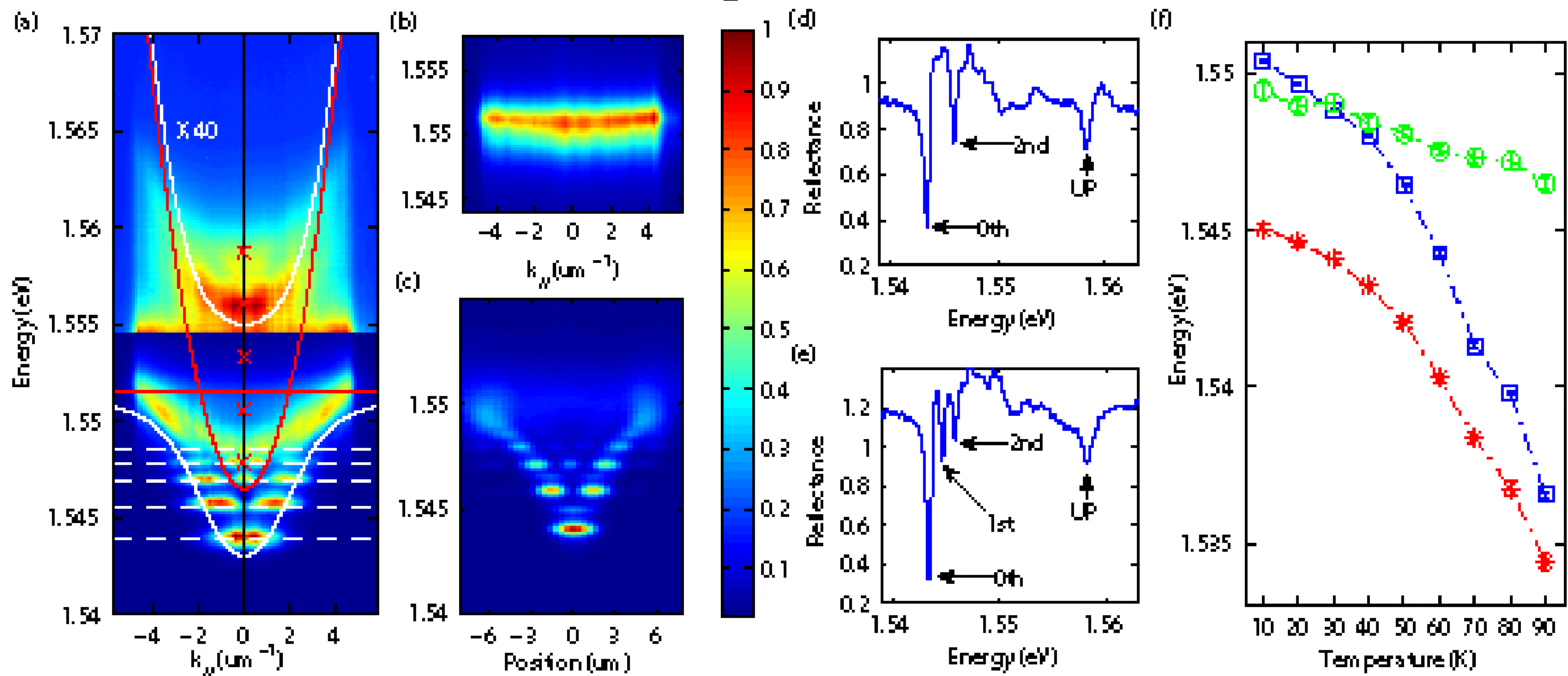

Figure 2

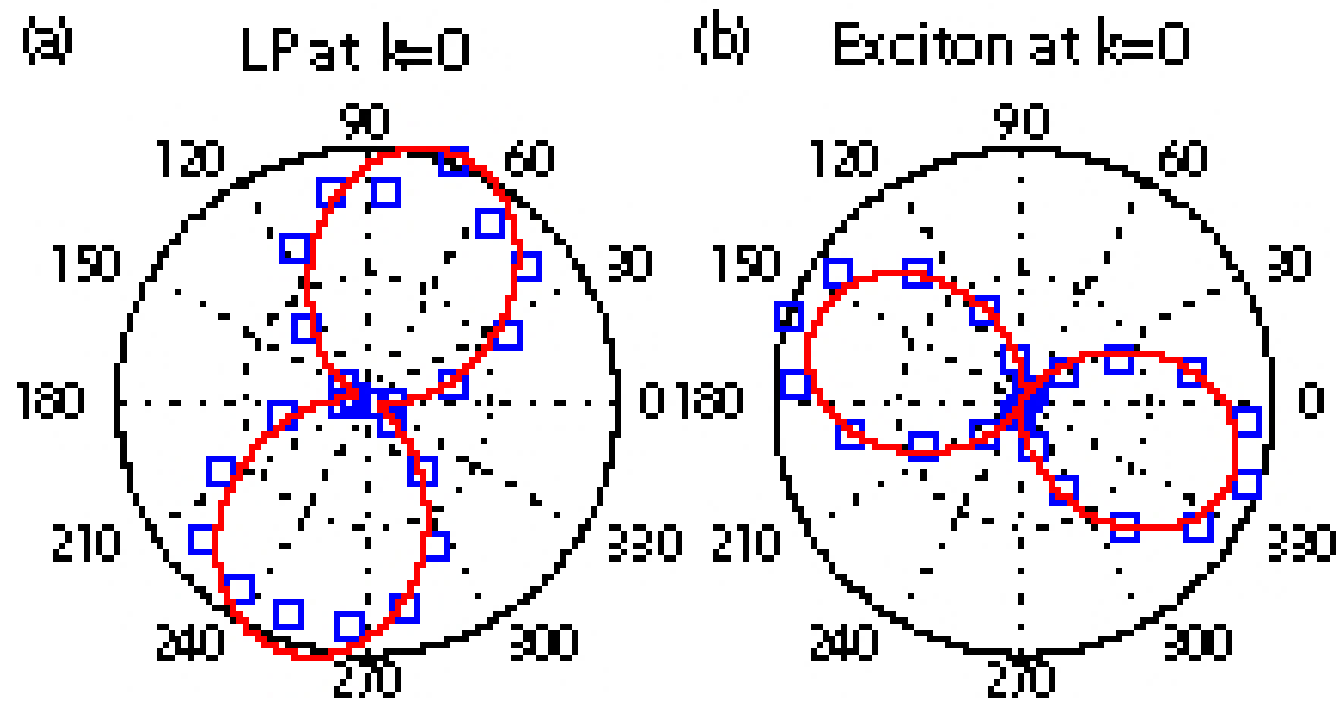

Figure 3

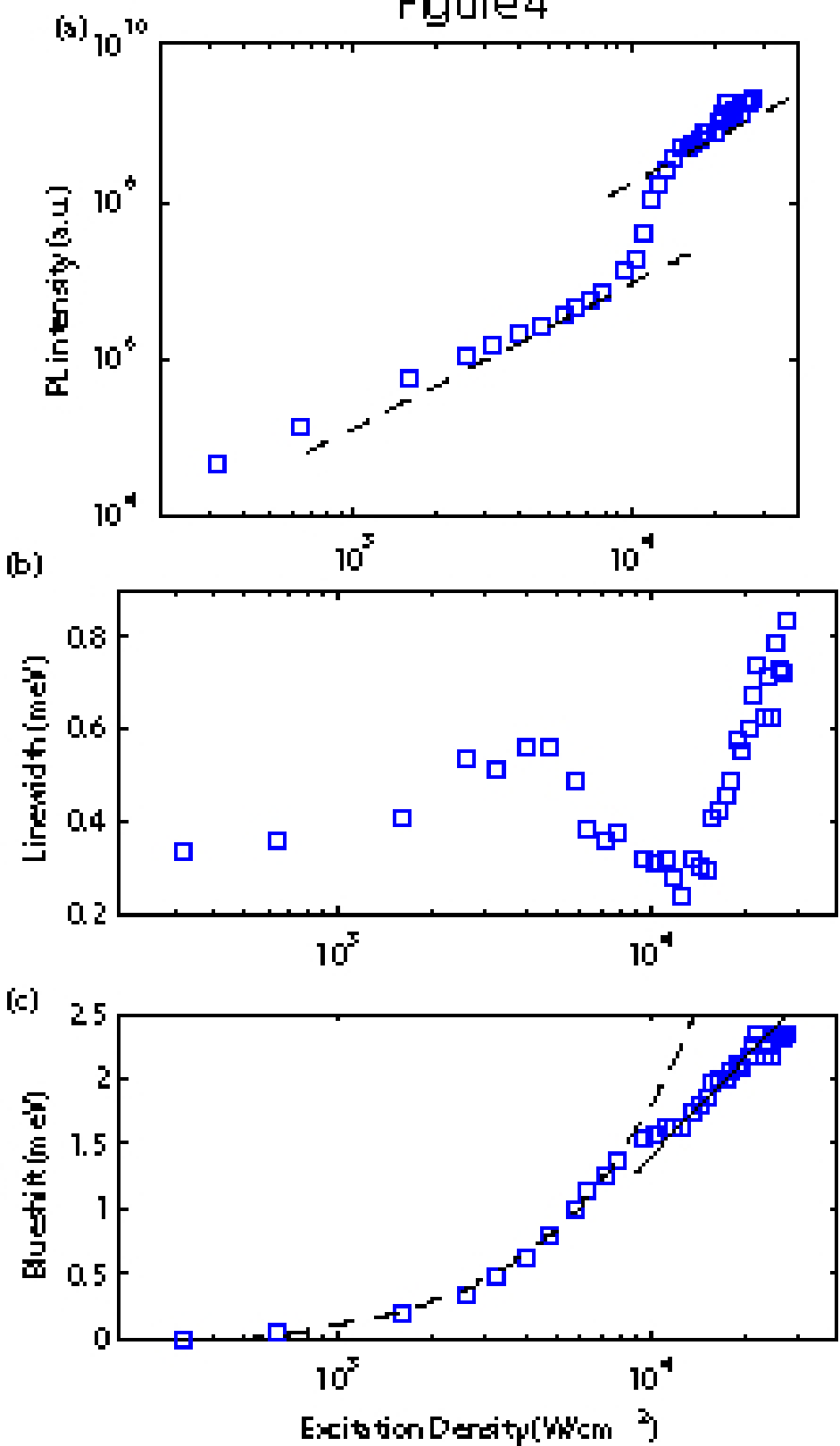

Figure 4